\documentclass[12pt]{article}

\usepackage{newtxtext,newtxmath}

\usepackage[normalem]{ulem}

\usepackage{xcolor}

\usepackage{graphicx}
\usepackage{amssymb}

\usepackage[letterpaper,margin=1in]{geometry}

\renewenvironment{abstract}
	{\quotation}
	{\endquotation}

\date{}

\makeatletter
\renewcommand{\fnum@figure}{\textbf{Figure \thefigure}}
\renewcommand{\fnum@table}{\textbf{Table \thetable}}
\makeatother

\usepackage{scicite}

\usepackage{url}
\usepackage{siunitx}

\def\scititle{
	Ultrafast Band-Gap Renormalization in Bilayer~Graphene
}
\title{\bfseries \boldmath \scititle}

\author{
	Eduard~Moos$^{1\ast}$,
    Zhi-Yuan~Deng$^{1}$,
    Hauke~Beyer$^{1}$,
    Arpit~Jain$^{4}$,
    Chengye~Dong$^{4, 5, 6}$,\and
    Li-Syuan~Lu$^{4}$,
    Joshua A.~Robinson$^{4, 5, 6, 7}$,
	Kai~Rossnagel$^{1,2,3}$,
    Michael~Bauer$^{1,3}$\and
	\small$^{1}$Institute of Experimental and Applied Physics, Christian Albrecht University of Kiel, \and
    \small 24118 Kiel, Germany.\and
    \small$^{2}$Ruprecht Haensel Laboratory, German Electron Synchrotron DESY, 22607 Hamburg, Germany.\and
    \small$^{3}$Kiel Nano, Surface and Interface Science KiNSIS, Christian Albrecht University of Kiel, \\
    \small 24118 Kiel, Germany\and
    \small$^{4}$Department of Materials Science and Engineering, The Pennsylvania State University, \and
    \small University Park, PA 16802, USA.\and
    \small$^{5}$2-Dimensional Crystal Consortium, Materials Research Institute, The Pennsylvania State University, \and
    \small University Park, PA 16802, USA.\and
    \small$^{6}$Center for 2-Dimensional and Layered Materials, The Pennsylvania State University, \and \small University Park, PA 16802, USA.\and
    \small$^{7}$Center for Atomically Thin Multifunctional Coatings, The Pennsylvania State University, \\
    \small University Park, PA 16802, USA.\and
	\small$^\ast$Corresponding author. Email: moos@physik.uni-kiel.de
}

\begin{document} 

\maketitle

\begin{abstract} \bfseries \boldmath
We demonstrate, by femtosecond time- and angle-resolved photoemission spectroscopy, that photoinduced interlayer charge transfer in a heterostructure consisting of Bernal-stacked bilayer graphene and a single atomic layer of silver on 6H-SiC(0001) transiently modulates the intrinsic potential landscape across the silver--graphene interface. This acts as an ultrafast optoelectronic gate that drives momentum-dependent band renormalizations, resulting in a transient band-gap opening on femtosecond timescales. Simultaneously, the photogenerated hot-carrier population enhances electronic screening, leading to subsequent closing of the band-gap beyond the thermal equilibrium value. These findings reveal two different mechanisms for photoinduced, reversible control of the electronic band structure in bilayer graphene---interlayer charge transfer and hot-carrier–enhanced screening---providing a general framework for the ultrafast control of electronic properties in graphene-based heterostructures. This opens up novel pathways for the realization of ultrafast optoelectronic devices and the exploration of correlated quantum phases in bilayer graphene under non-equilibrium conditions. 
\end{abstract}

\noindent

\newpage

Graphene-based materials exhibit remarkable optical and electronic properties \cite{NovoselovScience2004, ZhangNature2005, NairScience2008}, inspiring advances in next-generation nanoscale electronics \cite{ZhaiNatureReviewsElectricalEngineering2024}, and furthermore serve as archetypal systems for exploring novel electronic phenomena in low-dimensional confinement \cite{RozhkovPhysicsReports2016, AndreiNatureMaterials2020, PantaleonNatureReviewsPhysics2023}. These properties, including those governed by many-body interactions, can be substantially modified and functionalized via charge-transfer processes and band-gap engineering~\cite{CaoNature2018, MarchenkoScienceAdvances2018, SaitoNature2021, SeilerNature2022, SeilerNatureCommunications2024}.

Monolayer graphene (MLG), characterized by massless Dirac quasiparticles and intrinsically gapless, opens a band gap when a sublattice potential asymmetry is induced \cite{SemenoffPhysicalReviewLetters1984,NetoReviewsofModernPhysics2009, NiACSNano2008, ZhouNatureMaterials2007, BalogNatureMaterials2010, HuntScience2013, WoodsNaturePhysics2014}. Its control remains challenging and only partially reversible, thus limiting \textit{in operando} control and its suitability for device applications and fundamental research. In contrast, Bernal-stacked bilayer graphene (BLG) is particularly well suited for realizing a controllable band gap. By manipulating the interlayer potential asymmetry between the two hybridized graphene layers \cite{McCannPhysicalReviewLetters2006, McCannPhysicalReviewB2006, McCannReportsOnProgressInPhysics2013}---via substrate interactions \cite{ZhouNatureMaterials2007}, electrostatic gating \cite{CastroPhysicalReviewLetters2007, WangScience2008, ZhangNature2009, ZhangNatureCommunications2024}, or chemical doping \cite{OhteScience2006}---the semiconductor functionality can be controlled on electronic switching timescales. In particular, electrostatic gating provides highly precise and reversible control over both the interlayer potential asymmetry and charge-carrier density, enabling continuous tuning of the band gap. However, to the best of our knowledge, ultrafast, photoinduced control of the BLG band gap under non-equilibrium conditions on femtosecond timescales has not yet been experimentally demonstrated.

In this study, using time- and angle-resolved photoemission spectroscopy (tr-ARPES) \cite{RohwerNature2011}, we demonstrate that photoinduced interlayer charge transfer (ICT) \cite{HongNatureNanotechnology2014, AeschlimannScienceAdvances2020, FuScienceAdvances2021, ZhouScienceAdvances2021} in a BLG/monolayer silver (MLAg)/6H-SiC(0001) heterostructure can transiently modulate the intrinsic interlayer potential asymmetry in BLG, effectively acting as an optoelectronic gate that enables ultrafast control of the band gap. Simultaneously, the non-equilibrium distribution of hot carriers generated by photoexcitation transiently modulates electronic screening, further altering the band gap and providing an additional control mechanism. These results demonstrate that photoinduced ICT provides an ultrafast effective means to tailor the electronic and optoelectronic response of graphene-based van der Waals heterostructures, potentially enabling new opportunities for next-generation nanoscale electronics and for the study and manipulation of electrostatic gate--dependent phenomena on femtosecond timescales.

\subsection*{Electronic Structure of the BLG / MLAg / 6H-SiC(0001) Heterostructure}

\begin{figure}
	\centering
	\includegraphics[width=\textwidth]{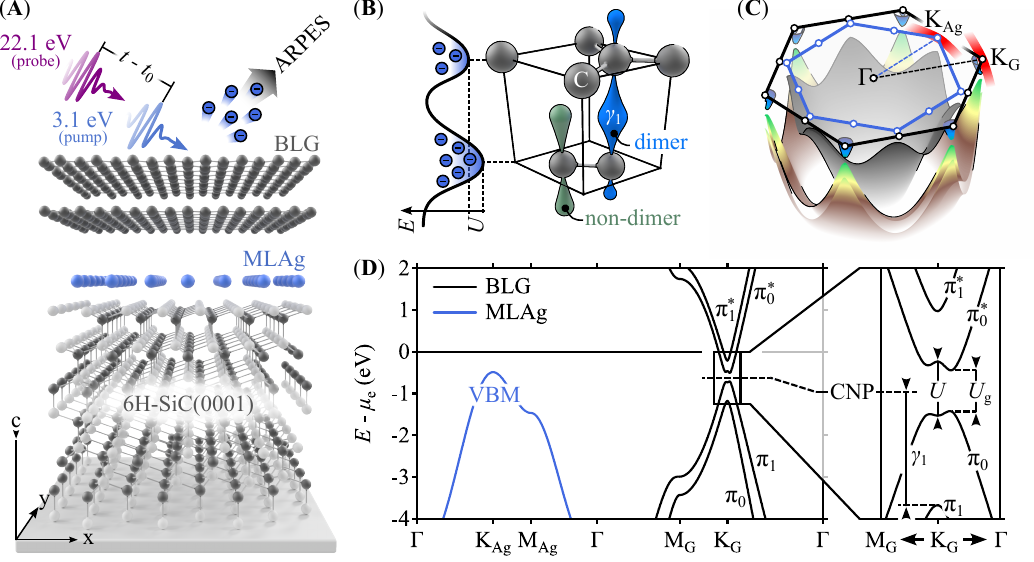}
	\caption{\textbf{Structural and electronic properties of Bernal-stacked BLG / MLAg / 6H-SiC(0001).}
		(\textbf{A}) Illustration of the tr-ARPES experiment used in this study and the crystalline structure of the sample with MLAg intercalated between 6H-SiC and BLG. The sample is photoexcited with pump pulses (\(\hbar\omega_\mathrm{pump} = \SI{3.1}{\electronvolt}\)) at an incident fluence of \(\SI{4.5}{\milli\joule\per\centi\meter\squared}\). Time-delayed probe pulses (\(\hbar\omega_\mathrm{probe} = \SI{22.1}{\electronvolt}\)) generate photoelectrons, which are detected in a hemispherical photoelectron analyzer. 
        (\textbf{B}) Unit cell of BLG: The bottom layer shows the two carbon atoms of the sublattices of a graphene layer and the associated \(p_z\) orbitals on dimer and non-dimer sites. The parameter \(\gamma_1\) reflects the hopping between \(p_z\) orbitals on dimer sites. The presence of an intrinsic \(U\) creates a gradient in the carrier density across the two layers. 
        (\textbf{C}) Hexagonal Brillouin zones and three-dimensional representation of the tight-binding valence band structures of BLG \cite{McCannPhysicalReviewLetters2006, McCannPhysicalReviewB2006, McCannReportsOnProgressInPhysics2013} and MLAg in the \(\mathrm{Ag}_{(1)}\) phase \cite{RosenzweigPhysicalReviewB2020}. The red lines indicate the directions along which the photoemission spectra were measured. 
        (\textbf{D}) Tight-binding model of the electronic band structures of BLG and MLAg in the \(\mathrm{Ag}_{(1)}\) phase along selected high-symmetry directions. The right panel shows a close-up of the BLG band structure near the CNP with the spectral signatures for \(\gamma_1\), \(U\) and \(U_\mathrm{g}\) indicated.}
	\label{fig:struct_electronic_prop}
\end{figure}

The heterostructure investigated in this study consists of semiconducting MLAg intercalated between an \textit{n}-type 6H-SiC(0001) substrate and epitaxial Bernal-stacked BLG (Fig.~\ref{fig:struct_electronic_prop}A) \cite{BriggsNatureMaterials2020, RosenzweigPhysicalReviewB2020, RosenzweigPhysicalReviewB2022, LeeNanoLetters2022, ArpitarXiv2025}. Chemisorbed on the substrate, MLAg in the \(\mathrm{Ag}_\mathrm{(2)}\) phase \cite{ArpitarXiv2025} adopts a high-order \((3\sqrt{3}\times 3\sqrt{3})R30^\circ\) surface structure in epitaxial registry with the underlying 6H-SiC surface, resulting in a supercell commensurate with the \((5 \times 5)\) 6H-SiC unit cell. In contrast, the BLG capping layer is physisorbed onto MLAg via van der Waals interactions, forming a \((6.25 \times 6.25)\) graphene supercell that arises from the commensurate alignment with the underlying MLAg surface. This configuration gives rise to two non-equivalent hexagonal Brillouin zones with distinct reciprocal lattice vectors, rotated by \(30^\circ\) relative to each other in reciprocal space (Fig.~\ref{fig:struct_electronic_prop}C).


Bernal-stacked BLG exhibits four \(\pi\)-bands, characterized at low energies by massive chiral quasiparticles with approximately parabolic dispersion near \(\mathrm{K}_\mathrm{G}\) in the BLG Brillouin zone (Fig.~\ref{fig:struct_electronic_prop}C, D) \cite{McCannPhysicalReviewLetters2006, McCannPhysicalReviewB2006, McCannReportsOnProgressInPhysics2013}. The low-energy \(\pi_0\) and \(\pi_0^\ast\) bands originate from \(p_z\) orbitals on non-dimer sublattice sites not involved in interlayer hybridization, whereas the high-energy \(\pi_1\) and \(\pi_1^\ast\) bands arise from strongly hybridized \(p_z\) orbitals on dimer sites. The energy splitting between the low- and high-energy \(\pi\)-bands directly reflects the contribution of the \(p_z\) orbitals located on dimer sites to the interlayer coupling, quantified by the vertical hopping parameter \(\gamma_1\) (Fig.~\ref{fig:struct_electronic_prop}B). As long as the inversion symmetry of the potential landscape between the two hybridized graphene layers is preserved, the low-energy \(\pi\)-bands remain gapless at \(\mathrm{K}_\mathrm{G}\). Breaking this inversion symmetry by introducing a finite interlayer potential asymmetry \(U\) (Fig.~\ref{fig:struct_electronic_prop}B) opens a band gap between the low-energy \(\pi\)-bands near the charge neutrality point (CNP) (Fig.~\ref{fig:struct_electronic_prop}D), driving BLG from a semimetallic to a semiconducting state. The BLG sample investigated here is intrinsically strongly \textit{n}-doped, with the CNP located \(\SI{600}{\milli\electronvolt}\) below the electrochemical potential \(\mu_\mathrm{e}\), yielding a doping-induced free carrier density of \(n_\mathrm{dop} \approx \SI{8.5e13}{\per\centi\meter\squared}\). This doping, combined with a pronounced gradient in bonding character across the heterostructure \cite{BriggsNatureMaterials2020}, gives rise to a strong interlayer potential asymmetry, resulting in \(U \approx \SI{300}{\milli\electronvolt}\) (Fig. \ref{fig:BLG_MLAg_SiC_sup_mat_BLG_band_gap}, Supplementary Materials), and a corresponding band gap \(U_\mathrm{g} \approx \SI{250}{\milli\electronvolt}\) near \(\mathrm{K}_\mathrm{G}\). 

Intercalated MLAg in the \(\mathrm{Ag}_\mathrm{(2)}\) phase between 6H-SiC and BLG exhibits a semiconducting electronic state resulting from substrate-induced breaking of inversion symmetry and the filling of silicon dangling bonds in 6H-SiC through donation of the Ag \(5s\) electrons \cite{WangarXiv2020}. The valence band structure is essentially identical to the \(\mathrm{Ag}_{(1)}\) phase and is characterized by a valence band maximum (VBM) at \(\mathrm{K}_\mathrm{Ag}\) of the MLAg Brillouin zone (Fig.~\ref{fig:struct_electronic_prop}C, D), with a binding energy of \(E_\mathrm{Ag}^\mathrm{K} \approx \SI{0.7}{\electronvolt}\) relative to \(\mu_\mathrm{e}\), and a saddle point at \(\mathrm{M}_\mathrm{Ag}\) with a binding energy of \(E_\mathrm{Ag}^\mathrm{M} \approx \SI{1.27}{\electronvolt}\) \cite{LeeNanoLetters2022, RosenzweigPhysicalReviewB2020, RosenzweigPhysicalReviewB2022, ArpitarXiv2025}, determined by fitting the corresponding energy distribution curves (EDCs) at \(\mathrm{K}_\mathrm{Ag}\) and \(\mathrm{M}_\mathrm{Ag}\) (Supplementary Materials).

\newpage

\subsection*{Interlayer Charge Transfer Dynamics}

\begin{figure}
	\centering
	\includegraphics[width=\textwidth]{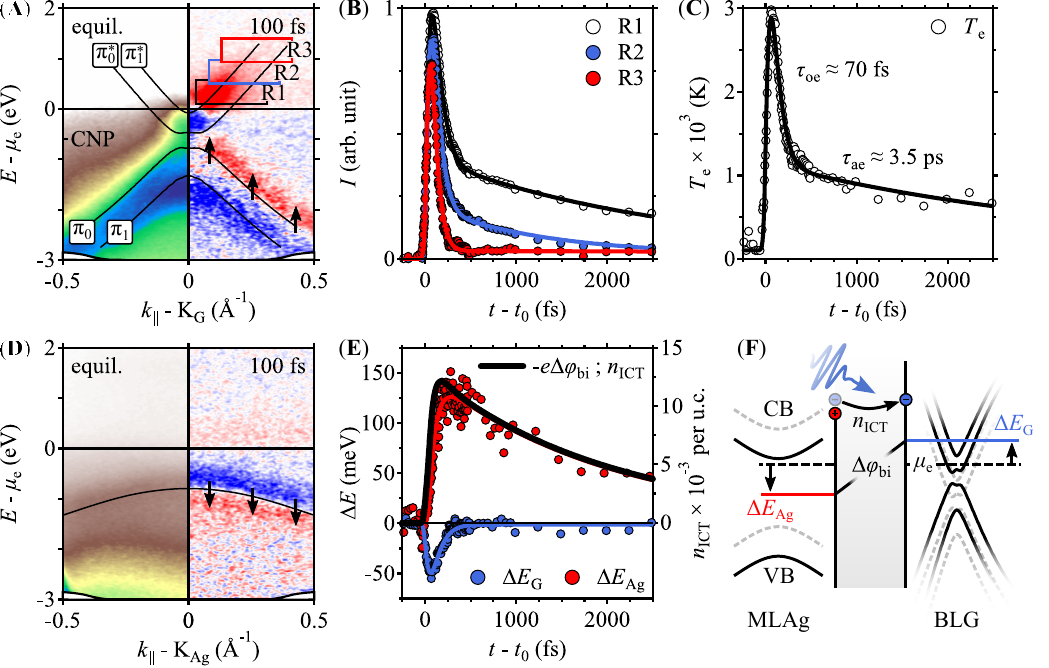} 
	\caption{\textbf{Non-equilibrium carrier and interlayer charge transfer dynamics.}
	(\textbf{A}) Equilibrium photoemission spectrum (left) of BLG at \(\mathrm{K}_\mathrm{G}\) and differential photoemission spectrum (right) at \(t - t_0 = \SI{100}{\femto\second}\), both superimposed with the equilibrium tight-binding band structure. The momentum cut is oriented perpendicular to the \(\mathrm{\Gamma}\)\text{--}\(\mathrm{K}_\mathrm{G}\) direction and is highlighted in red in Fig.\ref{fig:struct_electronic_prop}C, with \(k_\parallel\) denoting the electron wave vector parallel to the sample surface. The redistribution of spectral weight below \(\mu_\mathrm{e}\) indicates a rigid upshift of the \(\pi\) bands toward lower binding energies, as indicated by the arrows. Direction of the band shifts in panels A and D is referenced to the equilibrium \(\mu_\mathrm{e}\).
    (\textbf{B}) Integrated photoemission intensities from regions of interest R1--R3, indicated in panel A, as a function of pump--probe delay. 
    (\textbf{C}) Electronic temperature \(T_\mathrm{e}\) as a function of pump--probe delay. Solid lines in panels B and C represent multi-exponential fits to the experimental data (Supplementary Materials). The two time constants, \(\tau_\mathrm{oe}\) and \(\tau_\mathrm{ae}\), characterize the carrier--lattice thermalization process. 
    (\textbf{D}) Equilibrium photoemission spectrum (left) of MLAg at \(\mathrm{K}_\mathrm{Ag}\) and differential photoemission spectrum (right) at \(t - t_0 = \SI{100}{\femto\second}\). The momentum cut is oriented perpendicular to the \(\Gamma\)\text{--}\(K_\mathrm{Ag}\) direction and is highlighted in blue in Fig.~\ref{fig:struct_electronic_prop}C. Redistribution of spectral weight below \(\mu_\mathrm{e}\) indicates a rigid downshift of the valence band toward higher binding energies, as indicated by the arrows. 
    (\textbf{E}) Rigid band shifts \(\Delta E\) in BLG and MLAg and the change in the built-in potential across the MLAg--BLG interface, \(\Delta\varphi_\mathrm{bi}\), as a function of pump--probe delay. The scale on the right axis of the graph indicates the transferred charge density \(n_\mathrm{ICT}\), determined from \(\Delta\varphi_\mathrm{bi}\). The red and blue solid lines are exponential fits to the experimental data (Supplementary Materials). The black solid line, representing \(\Delta\varphi_\mathrm{bi}\), is obtained by subtracting the corresponding fits. 
    (\textbf{F}) Energy-band diagrams of MLAg (left) and BLG (right), illustrating the ICT-induced rigid band shifts in both BLG and MLAg, and the resulting \(\Delta\varphi_\mathrm{bi}\).}
	\label{fig:carrier_charge_transfer_dynamics}
\end{figure}

An angle-resolved photoemission spectrum of BLG measured in thermal equilibrium at \(\SI{100}{\kelvin}\) (Fig.~\ref{fig:carrier_charge_transfer_dynamics}A, left) reveals the characteristic low-energy band dispersion at \(\mathrm{K}_\mathrm{G}\). Following interband photoexcitation with \(\SI{35}{\femto\second}\) near-ultraviolet pump pulses, hot electrons populate the \(\pi_0^\ast\) and \(\pi_1^\ast\) bands above \(\mu_\mathrm{e}\). This population is evident in the differential photoemission spectrum (Fig.~\ref{fig:carrier_charge_transfer_dynamics}A, right), obtained by subtracting the equilibrium spectrum from the one measured at a pump--probe delay of \(t - t_0 = \SI{100}{\femto\second}\). In the color scale, blue indicates a loss and red a gain in spectral weight.

The transient hot-electron population dynamics (Fig.~\ref{fig:carrier_charge_transfer_dynamics}B) were determined by fitting the integrated spectral intensity within the selected energy–momentum regions of the photoemission spectrum indicated in the right panel of Fig.~\ref{fig:carrier_charge_transfer_dynamics}A (Supplementary Materials). The transient dynamics show a prompt rise immediately after photoexcitation, followed by a biexponential decay. A comparable behavior is observed in the transient electron temperature \(T_\mathrm{e}\) (Fig.~\ref{fig:carrier_charge_transfer_dynamics}C), determined by fitting transient EDCs with Fermi--Dirac functions (Supplementary Materials). The biexponential relaxation reveals two characteristic time constants, \(\tau_\mathrm{oe} \approx \SI{70}{\femto\second}\) and \(\tau_\mathrm{ae} \approx \SI{3.5}{\pico\second}\), corresponding to carrier--lattice thermalization via coupling to optical and acoustic phonons, respectively, in agreement with previous studies on graphitic materials \cite{KampfrathPhysicalReviewLetters2005, WangAppliedPhysicsLetters2010, JohannsenPhysicalReviewLetters2013}.

In addition to the hot-electron population dynamics in BLG, we observe pronounced transient rigid, i.e., momentum-independent, band shifts in both BLG and MLAg. The redistribution of spectral weight in BLG, observed below \(\mu_\mathrm{e}\) at \(t - t_0 = \SI{100}{\femto\second}\), indicates a rigid upward shift of the \(\pi\) bands toward lower binding energies. In contrast, for the valence band of MLAg at \(\mathrm{K}_\mathrm{Ag}\), the inverse behavior is observed, consistent with an opposite downward shift of the band toward higher binding energies (Fig.~\ref{fig:carrier_charge_transfer_dynamics}D). An analogous behavior is observed for the valence band at \(\mathrm{M}_\mathrm{Ag}\), as well as in the reference heterostructure consisting of monolayer graphene (MLG)/MLAg/6H-SiC(0001) (Fig.~\ref{fig:BLG_MLAg_SiC_sup_mat_MLG_MLAg}, Supplementary Materials).

Quantitative analysis at \(t - t_0 = \SI{100}{\femto\second}\), determined by fitting the corresponding EDCs (Supplementary Materials), reveals rigid band shifts of \(\Delta E_\mathrm{G} = -\SI{50}{\milli\electronvolt}\) in BLG and \(\Delta E_\mathrm{Ag} = +\SI{125}{\milli\electronvolt}\) in MLAg. These shifts provide direct evidence for photoinduced ICT between MLAg and BLG, dynamically modifying the interfacial dipole and its associated built-in potential \(\varphi_\mathrm{bi}\). Similar transient rigid band shifts arising from photoinduced ICT have previously been observed in tr-ARPES studies of other two-dimensional heterostructures \cite{AeschlimannScienceAdvances2020, BangeScienceAdvances2024, DongNatureCommunications2023}. 

The observed direction of the shifts unambiguously indicates a net negative charge transfer from MLAg to BLG, consistent with an accumulation of electrons in BLG and holes in MLAg (Fig.~\ref{fig:carrier_charge_transfer_dynamics}F). The transient dynamics of \(\Delta E_\mathrm{G}\) and \(\Delta E_\mathrm{Ag}\), along with the resulting change in the built-in potential, \(\Delta\varphi_\mathrm{bi} = (\Delta E_\mathrm{G} - \Delta E_\mathrm{Ag})/e\), where \(e\) denotes the elementary charge, are shown in Fig.~\ref{fig:carrier_charge_transfer_dynamics}E. The pronounced asymmetry in both amplitude and exponential decay is attributed to the strong \textit{n}-doping of BLG. The associated free-carrier density supports plasmon energies in the range of tens of millielectronvolts and corresponding screening times of several tens of femtoseconds \cite{HwangPhysicalReviewLetters2008, LvPhysicalReviewB2010, SensarmaPhysicalReviewB2010}. This enables efficient screening in BLG immediately after photoexcitation, causing the electrostatic potential to drop predominantly across MLAg---a behavior characteristic of semiconductor--metal interfaces \cite{BartolomeoPhysicsReports2016}. 

To estimate the transferred negative charge density \(n_\mathrm{ICT}\), the interface is modeled as a parallel-plate capacitor with a plate separation of \(d_\mathrm{AgG} \approx \SI{0.32}{\nano\meter}\), corresponding to the distance between the capacitively coupled BLG and MLAg layers \cite{ArpitarXiv2025}. Within this geometry, Gauss’s law yields the transferred negative charge density as \(n_\mathrm{ICT} = \varepsilon_0 \varepsilon_\mathrm{r,eff}\left|\Delta \varphi_\mathrm{bi}\right|/(ed_\mathrm{AgG})\). The value of the out-of-plane relative permittivity \(\varepsilon_\mathrm{r, eff} \approx 9.5\) is determined from the equilibrium BLG band structure, as discussed in detail in a subsequent section. In this model, the transient dynamics of \(n_\mathrm{ICT}\) follow \(\Delta\varphi_\mathrm{bi}\), reaching a maximum of approximately \(\num{1.3e-2}\,\mathrm{electrons\, per\,graphene\,unit\,cell}\) (\(\SI{2.5e13}{\per\centi\meter\squared}\)). The decay of \(n_\mathrm{ICT}\) on picosecond timescale reflects the back-transfer of charge carriers between the layers (Fig.~\ref{fig:carrier_charge_transfer_dynamics}E).

\textcolor{red}{The caption of Fig. \ref{fig:carrier_charge_transfer_dynamics} appears fragmented in the current manuscript version for the review due to excessive line spacing: (\textbf{E}) Rigid band shifts \(\Delta E\) in BLG and MLAg and the change in the built-in potential across the MLAg--BLG interface, \(\Delta\varphi_\mathrm{bi}\), as a function of pump--probe delay. The scale on the right axis of the graph indicates the transferred charge density \(n_\mathrm{ICT}\), determined from \(\Delta\varphi_\mathrm{bi}\). The red and blue solid lines are exponential fits to the experimental data (Supplementary Materials). The black solid line, representing \(\Delta\varphi_\mathrm{bi}\), is obtained by subtracting the corresponding fits. 
(\textbf{F}) Energy-band diagrams of MLAg (left) and BLG (right), illustrating the ICT-induced rigid band shifts in both BLG and MLAg, and the resulting \(\Delta\varphi_\mathrm{bi}\).}

\newpage

\subsection*{Ultrafast Band Renormalization}

\begin{figure}
	\centering
	\includegraphics[width=1\textwidth]{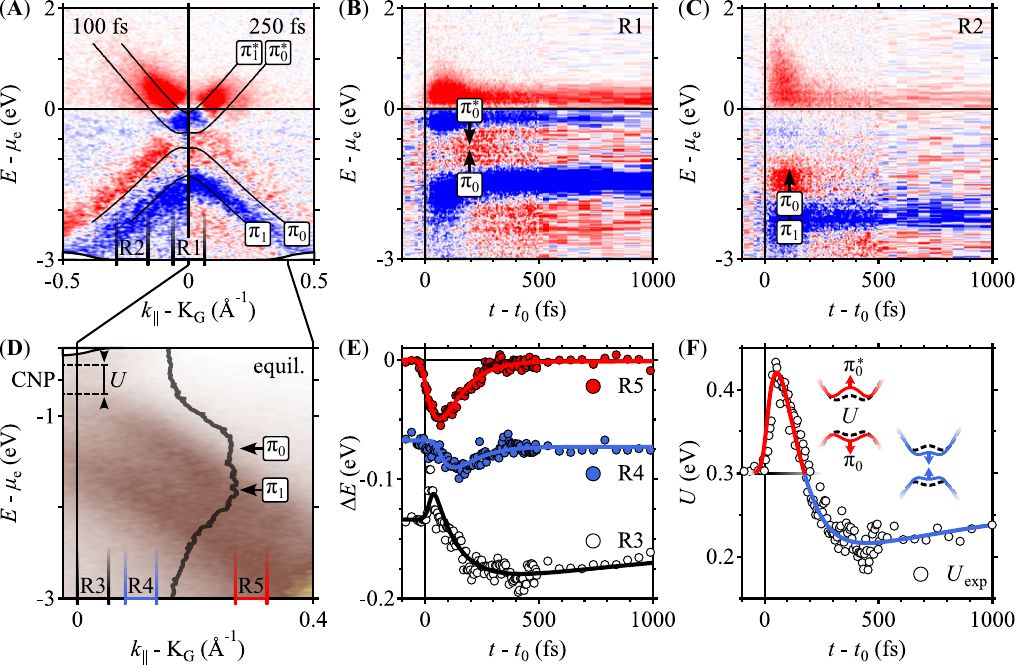}
	\caption{\textbf{Ultrafast band renormalization in BLG.} (\textbf{A}) Differential photoemission spectra at \(t - t_0 = \SI{100}{\femto\second}\) (left) and \(\SI{250}{\femto\second}\) (right) of BLG at \(\mathrm{K}_\mathrm{G}\). The comparison of the spectra reveals a delayed buildup of in-gap spectral weight between the \(\pi_0\) and \(\pi_0^\ast\) bands, as well as a delayed buildup of spectral weight below the \(\pi_1\) band. 
    (\textbf{B, C}) Transient differential EDCs for two distinct momentum regions of interest R1 (B) and R2 (C) marked in panel A. Relevant band energies are indicated. The arrows indicate the directions of the band renormalizations in panel B and the rigid band shift in panel C associated with the observed redistributions of spectral weight. 
    (\textbf{D}) Equilibrium photoemission spectrum measured with higher energy resolution (\(\Delta E_\mathrm{res}\approx \SI{190}{\milli\electronvolt}\)) compared to all other spectra shown. It enables to spectrally resolve the splitting of the \(\pi_0\) and \(\pi_1\) bands. The spectrum is superimposed with an EDC at \(k_\parallel - \mathrm{K}_\mathrm{G}\approx \SI{0.17}{\per\angstrom}\). The equilibrium interlayer potential asymmetry \(U\), which can be determined from the \(\pi_0\)--\(\pi_0^*\) splitting around CNP at \(\mathrm{K}_\mathrm{G}\) is indicated. (\textbf{E}) Changes in the \(\pi_0\) band binding energies \(\Delta E\) as a function of pump--probe delay for different momentum regions of interest R3--R5 indicated in panel D. The binding energies have been determined from fits to EDCs (Supplementary Materials). The changes in band energies for R3 and R4 have been vertically offset for clarity. (\textbf{F}) Interlayer potential asymmetry \(U\) as a function of pump--probe delay calculated by subtraction of the R5 transient from the R3 transient in panel E. Solid lines are guides to the eye. Here, red (blue) coloring indicates gap opening (closing) with respect to equilibrium as depicted by the insets.}
	\label{fig:fig_ultra_fast_bandrenormalization}
\end{figure}

In addition to the rigid band shift, the photoemission spectra of BLG reveal a transient, momentum-dependent band renormalization induced by photoexcitation. This observation is particularly evident between the \(\pi_0\) and \(\pi_0^\ast\) bands near \(\mathrm{K}_\mathrm{G}\), where at \(t - t_0 = \SI{100}{\femto\second}\), the differential photoemission spectrum deviates significantly from the behavior associated with the rigid band shift observed at off-\(\mathrm{K}_\mathrm{G}\) momenta (Fig.~\ref{fig:carrier_charge_transfer_dynamics}A, right panel, and Fig.~\ref{fig:fig_ultra_fast_bandrenormalization}A, left panel): At \(\mathrm{K}_\mathrm{G}\), the gain in spectral weight above the \(\pi_0\) and \(\pi_1\) bands, which is indicative of the rigid band shift, is virtually absent. Instead, a pronounced in-gap buildup of spectral weight emerges only at later pump--probe delays, as observed at \(t - t_0 = \SI{250}{\femto\second}\) (Fig.~\ref{fig:fig_ultra_fast_bandrenormalization}A, right). At this pump--probe delay, the redistribution of spectral weight associated with the rigid band shift has already decreased substantially.

Comparison of the transient EDCs at \(\mathrm{K}_\mathrm{G}\) (Fig.~\ref{fig:fig_ultra_fast_bandrenormalization}B) and at off-\(\mathrm{K}_\mathrm{G}\) momenta (Fig.~\ref{fig:fig_ultra_fast_bandrenormalization}C) emphasizes the distinct dynamics. In the latter case and consistent with the observed transient dynamics of the rigid band shift in BLG (Fig.~\ref{fig:carrier_charge_transfer_dynamics}E), a gain in spectral weight above the equilibrium band binding energy is observed immediately after photoexcitation and relaxes within approximately \(\SI{250}{\femto\second}\). In contrast, at \(\mathrm{K}_\mathrm{G}\), the in-gap buildup of spectral weight is significantly delayed and remains clearly observable at \(t - t_0 = \SI{500}{\femto\second}\). A further indication of band renormalization is the similarly delayed and long-lasting gain in spectral weight below the \(\pi_1\) band (Fig.~\ref{fig:fig_ultra_fast_bandrenormalization}A--C).

To quantify the band renormalization, we performed additional measurements under identical excitation conditions but with higher energy resolution \(\Delta E_\mathrm{res} \approx \SI{190}{\milli\electronvolt}\) (Fig.~\ref{fig:fig_ultra_fast_bandrenormalization}D). This enables us to spectrally resolve the splitting of the \(\pi_0\) and \(\pi_1\) bands and to independently assess the band renormalizations of the two sub-bands. The transient dynamics of the \(\pi_0\) band binding energies at representative momenta, corresponding to regions R3--R5 in Fig.~\ref{fig:fig_ultra_fast_bandrenormalization}D, are shown in Fig.~\ref{fig:fig_ultra_fast_bandrenormalization}E. For R3, which covers the band gap at \(\mathrm{K}_\mathrm{G}\), the band initially shifts toward higher binding energies, followed by a relaxation toward lower binding energies that even temporarily exceeds the equilibrium value for \(t - t_0 \gtrsim \SI{100}{\femto\second}\). In contrast, the transient dynamics at the off-\(\mathrm{K}_\mathrm{G}\) momenta covered by R5 predominantly follow the rigid band shift induced by ICT, with a continuous transition from R3 to R5 observed in the intermediate region R4.

Quantitative determination of the changes in the \(\pi_0^\ast\) and \(\pi_1^\ast\) band binding energies is not possible due to a significant redistribution of spectral weight near \(\mu_\mathrm{e}\), resulting from electron-hole generation and electron-gas heating following photoexcitation. However, the delayed gain in spectral weight below the equilibrium \(\pi_0^\ast\) band binding energy and \(\mu_\mathrm{e}\) for \(t - t_0 \gtrsim \SI{100}{\femto\second}\) (Fig. \ref{fig:fig_ultra_fast_bandrenormalization}A, B) indicates a shift toward higher binding energies, i.e., opposite to the transient behavior of the \(\pi_0\) band at \(\mathrm{K}_\mathrm{G}\). Such a behavior is consistent with a transient band-gap closing. This is further supported by the integrated spectral intensity in this region (Fig. \ref{fig:BLG_MLAg_SiC_Sup_Bandrenormalization}C, Supplementary Materials), where the transient behavior closely follows the \(\pi_0\) band shift (R3 in Fig.~\ref{fig:fig_ultra_fast_bandrenormalization}E).
We therefore interpret the delayed gain in spectral weight within the band-gap region, together with the associated \(\pi_0\) band shift toward lower binding energies for \(t - t_0 \gtrsim \SI{100}{\femto\second}\), as evidence of a transient reduction of \(U\) and closing of the band gap \(U_\mathrm{g}\), respectively.

In the band-gap region, the \(\pi_0\) band must also follow the global rigid band shift induced by the ICT. The observed change in band position therefore reflects a superposition of the band-gap dynamics and the rigid band shift, which can be disentangled by subtraction. The resulting transient behavior of \(U\) (Fig.~\ref{fig:fig_ultra_fast_bandrenormalization}F) is obtained from the \(\pi_0\) band shift at \(\mathrm{K}_\mathrm{G}\) and under the assumption of a fully anti-symmetric renormalization of the \(\pi_0\) and \(\pi_0^\ast\) bands. Following photoexcitation, \(U\) increases by approximately \(\SI{100}{\milli\electronvolt}\) within the first \(\SI{100}{\femto\second}\) and subsequently decreases within the following \(\SI{400}{\femto\second}\) to a value \(\SI{100}{\milli\electronvolt}\) below the equilibrium value of \(\SI{300}{\milli\electronvolt}\) observed before photoexcitation. The system then relaxes back to equilibrium with a characteristic time constant of approximately \(\SI{2}{\pico\second}\), a value that coincides with the slow component of carrier-lattice thermalization in BLG (Fig. \ref{fig:carrier_charge_transfer_dynamics}C). 

The quantitative analysis also reveals a renormalization of the \(\pi_1\) band that differs from the renormalization of the \(\pi_0\) band (Fig.~\ref{fig:blg_mlag_sic_sup_mat_interlayer_coupling}, Supplementary Materials). This indicates that, in addition to \(U\), the interlayer coupling \(\gamma_1\) between the two hybridized graphene layers is transiently modified upon photoexcitation. Overall, the observed band renormalizations are specific to BLG, where a breaking of the out-of-plane symmetry enables the formation and modification of an interlayer potential asymmetry. For monolayer graphene on MLAg/6H-SiC(0001), which lacks such structural degrees of freedom, only a rigid band shift indicative of ICT is observed, but no comparable band renormalization (Fig.~\ref{fig:BLG_MLAg_SiC_sup_mat_MLG_MLAg}B,C, Supplementary Materials).

\subsection*{Ultrafast Band-Gap Control}

Based on these observations, we formulate the central hypothesis of this study: Photoinduced ICT between MLAg/6H-SiC and BLG fundamentally modifies \(U\) and effectively acts as a single back gate, thereby controlling the single-particle band gap \(U_\mathrm{g}\) of BLG on ultrafast timescales.
 
\begin{figure}
	\centering
	\includegraphics[width = \textwidth]{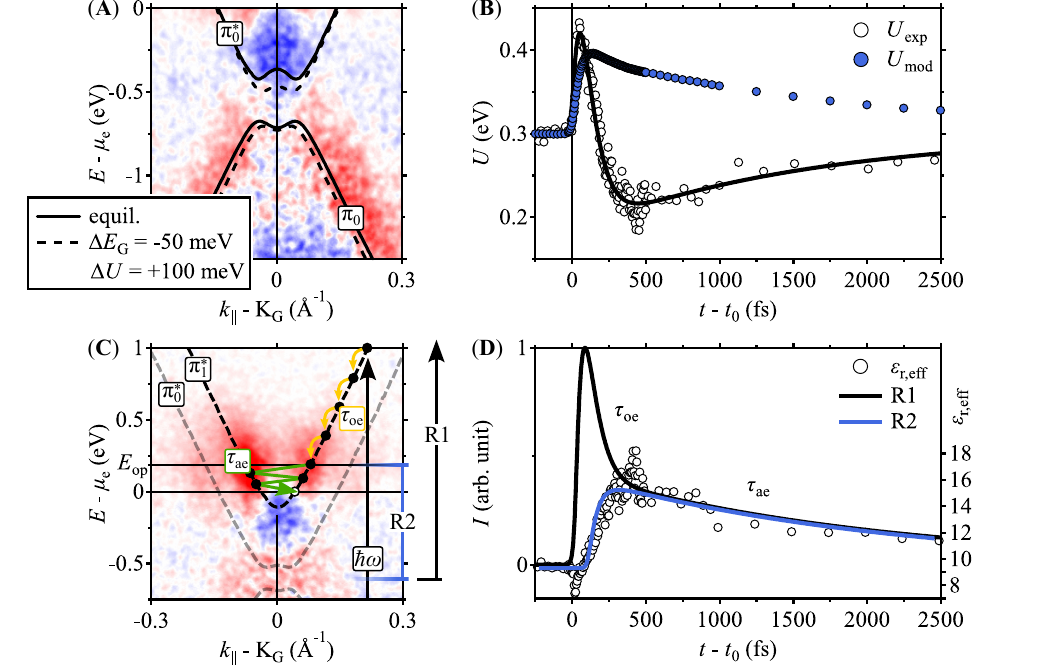} 
	\caption{\textbf{Single-back-gate model and non-equilibrium dielectric response of BLG.} (\textbf{A}) Tight-binding band structure of the \(\pi_0\) and \(\pi_0^\ast\) bands, calculated before photoexcitation (dashed line) and for an increased interlayer potential asymmetry of \(\Delta U = \SI{100}{\milli\electronvolt}\) (solid line) at \(t - t_0 = \SI{100}{\femto\second}\). The band structure for \(\Delta U = \SI{100}{\milli\electronvolt}\) was shifted by \(\Delta E_\mathrm{G} = -\SI{50}{\milli\electronvolt}\) to lower binding energies to account for the ICT-induced rigid band shift. For the \(\pi_0\) band near \(\mathrm{K}_\mathrm{G}\), the band shifts due to \(\Delta U\) and \(\Delta E_\mathrm{G}\) almost completely compensate each other, leading to a virtually absent spectral redistribution in the differential photoemission spectrum. 
    (\textbf{B}) Transient dynamics of \(U\) determined from the experiment (\(U_\mathrm{exp}\)) compared to the result of the single-back-gate model (\(U_\mathrm{mod}\)).  
    (\textbf{C}) Differential photoemission spectra at \(t - t_0 = \SI{100}{\femto\second}\) in the \(\pi^\ast\)-band region superimposed with the tight-binding band structure. The region of interest R2, marked by the blue line, includes the energy range above \(\mu_\mathrm{e}\), where carrier cooling is only possible by acoustic phonon emission (green arrow). The energy region R1 includes in addition the energy range above \(\mu_\mathrm{e}\), where carrier cooling is dominated by optical phonon emission (yellow arrow) \cite{YanPhysicalReviewB2008, MalicPhysicalReviewB2011}. The initial photoabsorption process using \(\hbar\omega = \SI{3.1}{eV}\) pump pulses (see black arrow) generates a nascent carrier distribution at \(E-\mu_\mathrm{e}\approx\SI{1.1}{eV}\). 
    (\textbf{D}) Comparison of the time dependence of the relative permittivity \(\varepsilon_\mathrm{r, eff}\) and the momentum-integrated photoemission intensities \(I\) in regions R1 and R2 marked in panel C. }
	\label{fig:BLG_tight_binding_and_screening}
\end{figure}

To substantiate this hypothesis theoretically, we first modeled the spectral redistribution observed in BLG within a Slonczewski--Weiss--McClure tight-binding framework \cite{McClurePhysicalReview1957, SlonczewskiPhysicalReview1958, McClurePhysicalReview1960}, extended with a self-consistent Hartree approximation following McCann \textit{et al.}~\cite{McCannPhysicalReviewB2006, McCannReportsOnProgressInPhysics2013}. Here, an ICT-induced change in \(U\) of BLG controls the band structure near \(\mathrm{K}_\mathrm{G}\), while the associated change in \(\varphi_\mathrm{bi}\) on the BLG side leads to a rigid band shift (Fig.~\ref{fig:BLG_tight_binding_and_screening}A). Both parameters were adjusted such that the calculations reproduce the experimentally observed maximum changes at \(t - t_0 = \SI{100}{\femto\second}\) in \(U\) (\(\Delta U \approx +\SI{100}{\milli\electronvolt}\)) and in the rigid band shift toward lower binding energies (\(\Delta E_\mathrm{G} \approx -\SI{50}{\milli\electronvolt}\)). The model accurately reproduces the opposing band shifts at \(\mathrm{K}_\mathrm{G}\), which are particularly evident from the virtually absent gain in spectral weight in the differential photoemission spectrum. For off-\(\mathrm{K}_\mathrm{G}\) momenta, and in agreement with the experimental observations, the model predicts only modest band shifts, predominantly resulting from the rigid band shift, while contributions from renormalization effects are negligible. This is further supported by the observation that the transient dynamics of the ICT-induced rigid band shift in BLG closely follows the band shift dynamics observed in MLG: MLG also shows an initial shift to lower binding energies and a subsequent relaxation within \(\SI{250}{\femto\second}\) (Fig.~\ref{fig:BLG_MLAg_SiC_sup_mat_MLG_MLAg}C, Supplementary Materials). As interlayer coupling is intrinsically absent in MLG, band renormalizations due to changes in \(U\) must be excluded here.

The transient dynamics of \(U\)---and accordingly \(U_\mathrm{g}\)---in response to the photoinduced ICT are described within a single-back-gate model following McCann \textit{et al.}~\cite{McCannReportsOnProgressInPhysics2013}, in which the MLAg/6H-SiC substrate acts as a time-varying back gate, analogous to an applied gate voltage or chemical doping. In this framework, the interlayer potential asymmetry is given by 

\begin{align}\label{eq:1}
    U\approx A\frac{n}{\varepsilon_\mathrm{r, eff}}\left[1-\frac{A}{2B}\frac{\gamma_1}{\varepsilon_\mathrm{r, eff}}\ln{\left(B\frac{\left|n\right|}{\gamma_1^2}\right)}\right]^{-1} \quad \text{with} \quad
    A=\frac{d_\mathrm{G}e^2}{2\varepsilon_0} \quad \text{and} \quad B=\pi\hbar^2v^2
\end{align}
and is determined by the hopping parameter \(\gamma_1\), the relative permittivity \(\varepsilon_\mathrm{r, eff}\), which accounts for the Hartree screening, and the total negative charge density \(n = n_\mathrm{dop} + n_\mathrm{ICT}\) in BLG, which accounts for both static doping \(n_\mathrm{dop}\) and transient charge transfer \(n_\mathrm{ICT}\). Further parameters are the BLG layer separation \(d_\mathrm{G} \approx \SI{0.34}{\nano\meter}\) \cite{McCannReportsOnProgressInPhysics2013, ArpitarXiv2025} and the band velocity \(v \approx \SI{0.87e6}{\meter\per\second}\), which was obtained by fitting the tight-binding model to the equilibrium photoemission spectra.

While \(\gamma_1\) and \(n\), as well as their transient dynamics, can be directly determined from the time-resolved photoemission spectra, \(\varepsilon_\mathrm{r,eff}\) is kept constant in our modeling. The value of \(\varepsilon_\mathrm{r,eff}\) was determined by treating it as an adjustable parameter in Eq.~\ref{eq:1} to reproduce the experimentally determined equilibrium \(U\), while taking \(\gamma_1\) and \(n\) into account. This yields the value of \(\varepsilon_\mathrm{r,eff} \approx 9.5\) already mentioned above, which is in good agreement with a value of \(\varepsilon_\mathrm{r,eff} \approx 7\) reported for the related system MLG/MLAu/6H-SiC(0001)~\cite{FortiNatureCommunications2020}.

Fig.~\ref{fig:BLG_tight_binding_and_screening}B compares the transient dynamics of the experimentally measured and modeled \(U\). The model well reproduces the initial increase in \(U\) by about \SI{100}{\milli\electronvolt} observed in the experiment and the corresponding band-gap opening. The results imply a linear dependence of \(U\) on \(n_\mathrm{ICT}\) with  \( \mathrm{d}U/\mathrm{d}n_{\mathrm{ICT}} \approx \SI{4e-12}{\milli\electronvolt\per\centi\meter\squared} \). This value corresponds reasonably well to the value of approximately \(\SI{7e-12}{\milli\electronvolt\per\centi\meter\squared}\) extracted from the data of a static study by Ohta \textit{et al.} on band-gap tuning in BLG~\cite{OhteScience2006}, in which the charge transfer was induced by chemical doping via potassium adsorption. The difference in \( \mathrm{d}U/\mathrm{d}n_{\mathrm{ICT}}\) can be explained by differences in doping level and the dielectric environment \cite{GavaPhysicalReviewB2009}. In contrast to the initial dynamics, model and experiment deviate significantly from each other for \(t - t_0 \ge \SI{200}{\femto\second}\). We attribute these differences to excited charge carriers generated by interband photoexcitation within BLG, which are not taken into account in the model. Interband photoexcitation results in an increase in the free carrier density, which generally alters the screening capabilities and \(\varepsilon_\mathrm{r, eff}\) of graphene-based material, respectively \cite{TielrooijNaturePhysics2013, TomadinScienceAdvances2018, AgarwalNaturePhotonics2023}. 
Importantly, due to the high pump fluence (\(\SI{4.5}{\milli\joule\per\centi\meter\squared}\)) the experiments are carried out in the saturation absorption regime for BLG and, due to the high pump photon energy (\(\SI{3.1}{\electronvolt}\)), the interband transitions couple to a high density of electronic states. A rough estimate based on reference data from MLG \cite{WinzerAppliedPhysicsLetters2012, MaloufAppliedPhysicsLetters2019} yields \(n_\mathrm{eh} \approx \SI{1e14}{\per\centi\meter\squared}\), comparable to the density of doping-induced charge carriers \(n_\mathrm{dop} \approx \SI{8.5e13}{\per\centi\meter\squared}\). In contrast to the increase in free carrier density, the (significant) increase in electronic temperature due to photoexcitation has only a negligible effect on the screening properties of BLG \cite{LvPhysicalReviewB2010}.

The differences between experimental observations and the single-back-gate model can thus serve as a probe for the time-dependent behavior of \(\varepsilon_\mathrm{r, eff}\). The result of this comparison is shown in Fig.~\ref{fig:BLG_tight_binding_and_screening}D. The value of \(\varepsilon_\mathrm{r,eff}\) was determined for each pump--probe delay using Eq.~\ref{eq:1}, with the experimental values of \(U\), \(\gamma_1\), and \(n\) being used for the calculation. Approximately \(\SI{100}{\femto\second}\) after photoexcitation, \(\varepsilon_\mathrm{r, eff}\) begins to increase significantly, reaches a maximum at  \(t - t_0 \approx \SI{400}{\femto\second}\), and relaxes back to its equilibrium value with a characteristic time constant of \(\tau \approx \SI{2}{\pico\second}\).

The temporal evolution of \(\varepsilon_\mathrm{r,eff}\) (Fig.~\ref{fig:BLG_tight_binding_and_screening}D) is superimposed with integrated photoemission intensities from two selected regions of interest (R1 and R2 in Fig.~\ref{fig:BLG_tight_binding_and_screening}C) that cover part of the \(\pi^\ast\) bands. The energy range covered by R1 extends beyond the resonant excitation energy in the photoabsorption process. Its integrated photoemission intensity serves as a measure of the total free-carrier density in BLG at each pump--probe delay. Remarkably, the transient dynamics of \(\varepsilon_\mathrm{r, eff}\) shows a significantly delayed response and only matches the total charge-carrier density after a maximum at \(t - t_0 = \SI{400}{\femto\second}\). This maximum coincides with the signature for the temporal crossover from optical to acoustic phonon-mediated carrier cooling, characterized by the time constants \(\tau_\mathrm{oe}\) and \(\tau_\mathrm{ae}\).

Better agreement is observed when compared to the integrated photoemission intensity from R2, which covers the energy range up to an excess energy of \(\sim\SI{160}{\milli\electronvolt}\) above \(\mu_\mathrm{e}\), the optical phonon threshold for electron-optical phonon interaction \cite{YanPhysicalReviewB2008}. This observation indicates that low-energy charge carriers near \(\mu_\mathrm{e}\) screen most effectively. This is consistent with the prediction that electronic states near \(\mu_\mathrm{e}\) contribute most to the polarizability in BLG, and therefore its screening capability~\cite{HwangPhysicalReviewLetters2008, LvPhysicalReviewB2010, SensarmaPhysicalReviewB2010}. Photoexcited electrons relax into these low-energy states via carrier-lattice cooling, contributing to a delayed screening in line with other experimental observations~\cite{TielrooijNaturePhysics2013, TomadinScienceAdvances2018, AgarwalNaturePhotonics2023}.

Overall, this leads to the following scenario for the observed ultrafast band-gap dynamics of BLG in the investigated heterosystem: The initial band-gap opening within the first \(\SI{100}{\femto\second}\) is the result of a photoinduced net negative charge transfer from MLAg to BLG. However, the delayed accumulation of additional photogenerated free carriers in BLG near \(\mu_\mathrm{e}\) enhances electronic screening and counteracts the band-gap opening. In the following, screening becomes so effective that the band gap closes beyond its thermodynamic equilibrium value. As the photogenerated charge-carrier density decays via interband recombination, electronic screening is reduced, and the band gap opens until it reaches its equilibrium value again.

\subsection*{Conclusion}

Using femtosecond tr-ARPES, two counteracting mechanisms for ultrafast band gap control in BLG were identified. First, a photoinduced net negative interlayer charge transfer from MLAg to BLG in a BLG/MLAg/6H-SiC(0001) heterostructure causes a significant opening of the band gap: Photoexcitation with \(\SI{400}{\nano\meter}, \SI{35}{\femto\second}\) laser pulses at excitation fluences of several \(\SI{}{\milli\joule\per\centi\meter\squared}\) increases the band gap by approximately \(\SI{100}{\milli\electronvolt}\). Second, the delayed build-up of electronic screening due to the photoexcitation of charge carriers within BLG closes the band gap. Here, we observe changes of the band gap of up to approximately \(\SI{200}{\milli\electronvolt}\).   

These results show that ultrafast optical excitation of BLG, embedded in suitable heterostructures, can effectively function as a single optoelectronic back gate, enabling reversible band-gap control in BLG on the femtosecond timescale. The detailed response depends on the electrostatic preset of the heterosystem and the charge carrier densities injected into and photoexcited within BLG. One of the parameters that allows the system to be tailored to the specific requirements is the neighboring heterostructure environment. Its dielectric properties determine the electrostatic preset. In addition, its band structure and electronic coupling to BLG influence the efficiency and direction of the photoinduced charge transfer, allowing the amplitude and direction of the band gap response to be tuned. Single layers of semiconducting transition metal dichalcogenides (TMDCs) embedded in such heterostructures are particularly well suited because of their strong absorption near excitonic resonances and valley-selective excitation via optical helicity \cite{SchneiderNatureCommunications2018}, while high charge transfer rates have also been demonstrated for such systems in interaction with MLG \cite{AeschlimannScienceAdvances2020, FuScienceAdvances2021, ZhouScienceAdvances2021}.

The band-gap closing due to the excited charge carrier screening can be tuned in amplitude and timing by the wavelength of the excitation laser source. Rapid closing of the band gap can be achieved, for example, by tuning the excitation wavelength close to resonance with \(\mu_e\), thereby eliminating the delay due to charge carrier relaxation. At even longer wavelengths, the photoexcitation of charge carriers within BLG and the associated band gap closing mechanisms could be completely suppressed. In this case, band-gap control would be achieved exclusively by ICT, if it is supported by the electronic structure of the substrate at these wavelengths. Overall, the presented results provide new insight into the interplay between charge transfer and ultrafast screening in layered 2D-materials and open pathways for light-driven engineering of electronic properties in graphene-based ultrafast optoelectronic devices.


\clearpage 

%
\bibliography{ref} 

\bibliographystyle{sciencemag}

%
%
%
%
%
%


\section*{Acknowledgments}

\paragraph*{Funding:}

E.M. and M.B. were supported by the German Research Foundation (DFG), Project ID 465690255. A.J., C.D., L.L., and J.A.R. were supported by the National Science Foundation (NSF) through Award No. DMR-2011839 (Penn State MRSEC—Center for Nanoscale Science). C.D. and J.A.R. additionally supported by the Two-Dimensional Crystal Consortium–Materials Innovation Platform (2DCC-MIP) under NSF Cooperative Agreement No. DMR-2039351.

\paragraph*{Author contributions:}

E.M. and M.B. conceived the experiments. E.M., Z.D., and H.B. carried out tr-ARPES measurements and performed the data analysis and the calculations. C.D. and L.L. synthesized the epitaxial graphene samples. A.J. performed silver intercalation and sample characterization.  E.M. and M.B. co-wrote the paper. E.M., A.J., C.D., and J.A.R., K.R., and M.B. contributed to the scientific discussions.    
\paragraph*{Competing interests:}

There are no competing interests to declare.

\paragraph*{Data and materials availability:}

All data needed to evaluate the conclusions in the paper are present in the paper and/or the Supplementary materials. All raw and processed data are publicly available on Zenodo (DOI: 10.5281/zenodo.18229471).


\subsection*{Supplementary materials}
Materials and Methods\\
Supplementary Text\\
Figs. S1 to S4\\


\newpage


\renewcommand{\thefigure}{S\arabic{figure}}
\renewcommand{\thetable}{S\arabic{table}}
\renewcommand{\theequation}{S\arabic{equation}}
\renewcommand{\thepage}{S\arabic{page}}
\setcounter{figure}{0}
\setcounter{table}{0}
\setcounter{equation}{0}
\setcounter{page}{1} 


\begin{center}
\section*{Supplementary Materials for\\ \scititle}

	Eduard~Moos$^{1\ast}$,
    Zhi-Yuan~Deng$^{1}$,
    Hauke~Beyer$^{1}$,
    Arpit~Jain$^{4}$,
    Chengye~Dong$^{4, 5, 6}$,\and
    Li-Syuan~Lu$^{4}$,
    Joshua A.~Robinson$^{4, 5, 6, 7}$,
	Kai~Rossnagel$^{1,2,3}$,
    Michael~Bauer$^{1,3}$\and

\small$^\ast$Corresponding author. Email: moos@physik.uni-kiel.de
\end{center}

\subsubsection*{This PDF file includes:}
Materials and Methods\\
Supplementary Text\\
Figures S1 to S4\\

\newpage

\subsection*{Supplementary Information}

\subsubsection*{Experimental Method}

Time- and angle-resolved photoemission spectroscopy (tr-ARPES) is used to investigate the ultrafast non-equilibrium dynamics of charge carriers and band renormalization. The sample is photoexcited with a \(\SI{35}{\femto\second}\) near-ultraviolet pump pulse (\(\hbar\omega_\mathrm{pump} = \SI{3.1}{\electronvolt}\)) with an incident fluence of \(\SI{4.5}{\milli\joule\per\centi\meter\squared}\), driving the electronic system into a non-equilibrium state. Extreme-ultraviolet probe pulses (\(\hbar\omega_\mathrm{probe} = \SI{22.1}{\electronvolt}\)) are generated in a high-harmonic generation source and are used to map transient changes in the electronic structure via photoemission. Photoelectrons are detected using a hemispherical electron analyzer. For most experiments the energy resolution \(\Delta E_\mathrm{res}\) was set to \(\SI{250}{\milli\electronvolt}\), for selected experiments, a higher energy resolution of \(\SI{190}{\milli\electronvolt}\) was chosen. The time resolution of the tr-ARPES experiment is \(\Delta t_\mathrm{res} \approx \SI{35}{\femto\second}\) corresponding to the FWHM of the pump–probe cross-correlation. All measurements are performed under ultra-high vacuum at \(\SI{100}{\kelvin}\), with Cu/Ta contacts used to ground the sample and suppress charging effects. Further details on the experimental setup are described in Refs. \cite{RohwerNature2011, EichJournalofElectronSpectroscopyandRelatedPhenomena2014}.

\subsubsection*{Zero-layer Graphene and Epitaxial Graphene Synthesis}

Zero-layer graphene (ZLG) and epitaxial graphene (EG) were synthesized on the Si-terminated (0001) face of semi-insulating 6H-SiC substrates (Coherent Corp.) via thermal silicon sublimation. The substrates were initially annealed at \(\SI{1400}{\degreeCelsius}\) for \(\SI{30}{\minute}\) in a \(\SI{10}{\percent}\)~\(\mathrm{H_2 / Ar}\) atmosphere at \(700\,\mathrm{Torr}\) to remove polishing damage and surface contaminants. ZLG formation was achieved by subsequent annealing at \(\SI{1600}{\degreeCelsius}\) in \(700\,\mathrm{Torr}\) of pure Ar for \(\SI{30}{\minute}\). Undergrown EG, characterized by partial ZLG coverage, was obtained by systematically tuning the annealing temperature (\(\num{1600}\)--\(\SI{1650}{\degreeCelsius}\)) and duration (\(\num{15}\)–\(\SI{30}{\minute}\)), enabling controlled formation of zero and monolayer graphene domains.

\subsubsection*{2D Silver Intercalation}

Silver intercalation was carried out using a one-inch outer-diameter quartz tube furnace (Thermo Scientific Lindberg/Blue M Mini-Mite). Undergrown EG/SiC substrates (approximately \(1\,\times\,\SI{0.5}{\centi\meter\squared}\)) were placed face-down in an alumina crucible positioned above \(\SI{80}{\milli\gram}\) of high-purity silver powder (\(\SI{99.99}{\percent}\), Sigma-Aldrich). The furnace tube was evacuated to approximately \(5\,\mathrm{mTorr}\), leak-checked, and backfilled with argon to \(500\,\mathrm{Torr}\). The system was ramped to \(\SI{900}{\degreeCelsius}\) at \(\SI{50}{\degreeCelsius\per\minute}\) under continuous argon flow (\(50\,\mathrm{sccm}\)), held at temperature for \(\SI{70}{\minute}\), and subsequently cooled to room temperature under an inert atmosphere. This process resulted in the formation of a monolayer \(\mathrm{Ag_{(2)}}\) phase intercalated beneath mono- and bilayer graphene. Additional sample synthesis and characterization details are provided in previous work \cite{ArpitarXiv2025}. 

\subsubsection*{Sample Cleaning}

The cleaning of the heterostructure was performed directly under high-vacuum conditions using laser annealing. Near-ultraviolet laser pulses (\(\SI{3.1}{\electronvolt}\)) with a pulse duration of \(\SI{35}{\femto\second}\) were employed. The incident fluence was approximately \(\SI{40}{\milli\joule\per\centi\meter\squared}\). During the procedure, the sample was raster-scanned with a lateral step size of \(\SI{50}{\micro\meter}\), and each position was irradiated for \(\SI{10}{\second}\).

\subsubsection*{Determination of the Doping-induced Equilibrium Charge-Carrier Density in BLG}

The electronic band structure and the density of electronic states (DOS) of BLG were calculated within the Slonczewski--Weiss--McClure tight-binding framework \cite{McClurePhysicalReview1957, SlonczewskiPhysicalReview1958, McClurePhysicalReview1960}. To obtain a numerical representation of the DOS, the tight-binding model was fitted to the measured equilibrium photoemission spectra. The corresponding band energies \(E(k_x, k_y)\) of the \(\pi\) bands, were determined on a dense, uniform \(k\)-point grid spanning the full BLG Brillouin zone. The energy axis was discretized into small intervals of width \(\Delta E\). For each energy interval \(i\), the number of electronic states within the interval \([E_i, E_i + \Delta E]\) was counted as

\begin{align}
    N_i = \sum_{n,k_x,k_y} \Theta(E_n(k_x, k_y) \in [E_i, E_i + \Delta E]),
\end{align}

where \(\Theta\) is the step function, equal to \(1\) if the condition is satisfied and \(0\) otherwise. Spin and valley degeneracies were taken into account where appropriate. The DOS was then obtained by normalizing the counts by the total number of \(k\)-points \(N_k\) and the bin width \(\Delta E\), i.e., 

\begin{align}
    \mathrm{DOS}(E_i) = \frac{N_i}{N_k \, \Delta E}.
\end{align}

This histogram-based approach yields a discrete representation of the DOS. The resulting DOS was then combined with the equilibrium Fermi--Dirac function \(\mathrm{FD}(E; \mu_\mathrm{e}, T_\mathrm{e})\) \and fitted to the experimental energy density of states, taking into account the finite energy resolution of the experiment by convolution with a Gaussian function \(\mathrm{G}(E; \Delta E_\mathrm{res})\). The doping-induced equilibrium charge-carrier density \(n_\mathrm{dop}\) is then given by:

\begin{align}
    n_\mathrm{dop} &= \int_{-\infty}^{\infty} \mathrm{DOS}(E) \, \left[\mathrm{FD}(E; \mu_\mathrm{e}, T_\mathrm{e}) \ast \mathrm{G}(E; \Delta E_\mathrm{res})\right] \, \mathrm{d}E \
\end{align}
with
\begin{align}
    \mathrm{FD}(E; \mu_\mathrm{e}, T_\mathrm{e}) &=  \frac{1}{\exp\left(\frac{E - \mu_\mathrm{e}}{k_\mathrm{B} T_\mathrm{e}}\right) + 1}.
    \label{eq:fermi_dirac}
\end{align}

Here, \(\mu_\mathrm{e}\) is the electrochemical potential and \(T_\mathrm{e}\) the electronic temperature. This results in a charge-carrier density of \(n_\mathrm{dop} \approx \SI{0.85e14}{\per\centi\meter\squared}\), in good agreement with Luttinger’s theorem for the reference system MLG/MLAg/6H-SiC(0001) in the \(\mathrm{Ag}_\mathrm{(2)}\) phase \cite{ArpitarXiv2025}.

\subsubsection*{Fitting Function for the Analysis of the Data Shown in Figs. 2(A), 2(E), 3(E), and 4(D)}

For each pump--probe delay integrated spectral intensities \(I\)  were determined from the regions of interest marked in the photoemission spectrum at each pump--probe delay. The resulting transient intensity was fitted using the following piecewise function:

\begin{align}
f_\mathrm{fit}(t) &= f(t) * \mathrm{G}(t; \Delta t_\mathrm{res}), \\
f(t) &=
\begin{cases}
f_0, & t < t_0 \\
f_0 + \sum_i A_i \exp\left(-\frac{t - t_0}{\tau_i}\right), & t \ge t_0
\end{cases}
\label{eq:transient_dynamics_fitting_procedure}
\end{align}

\(f(t)\) accounts for the constant intensity \(f_0\) prior to photoexcitation (\(t < t_0\)) and a multi-exponential response after photoexcitation (\(t \ge t_0\)). The convolution with a Gaussian function \(\mathrm{G}(t; \Delta t_\mathrm{res})\) accounts for the finite temporal resolution of the experiment. The sum over \(i\) channels allows for multiple contributions with characteristic time constants \(\tau_i\) and amplitudes \(A_i\). An analog fitting procedure was used to analyze the observed transient band shifts and the evaluated transient for \(\varepsilon_\mathrm{r,eff}\). 

\subsubsection*{Determination of the Electron Temperature in BLG}

To determine the transient electron temperature \(T_\mathrm{e}\) in BLG, the photoemission spectra were first integrated over the entire momentum field of view. After subtraction of a Shirley background, which accounts for inelastic scattering during the photoemission process, the resulting energy distribution curves (EDCs) were fitted near \(\mu_\mathrm{e}\) with a Fermi--Dirac function \(\mathrm{FD}(E; \mu_\mathrm{e}, T_\mathrm{e})\) (Eq. \ref{eq:fermi_dirac}). The fitting function was convolved with a Gaussian function to account for the finite energy resolution of the experiment. This fitting procedure was applied to the experimental EDCs for each pump--probe delay. The transient \(T_\mathrm{e}\) was then fitted with a fitting procedure analogous to Eq. \ref{eq:transient_dynamics_fitting_procedure} to account for the sudden rise in \(T_\mathrm{e}\) after photoexcitation and the bi-exponential cooling behaviour due to coupling to the phonon bath. The fitting yielded the characteristic time constants \(\tau_\mathrm{oe}\) and \(\tau_\mathrm{ae}\).

\subsubsection*{Determination of the Band Binding Energies in BLG and MLAg}

To determine the band binding energies of the BLG \(\pi_0\) and \(\pi_1\) bands, as well as the valence band in MLAg, the EDCs in the regions of interest were first corrected by subtracting a Shirley background to account for inelastic scattering during the photoemission process. Each EDC exhibits well-defined peaks, which were fitted using Voigt functions \(V(E) = G(E; \Delta E_\mathrm{res}) \ast L(E)\), a convolution of a Gaussian function \(G(E; \Delta E_\mathrm{res})\) with a Lorentzian function \(L(E)\). Here, the Gaussian contribution represents the energy resolution \(\Delta E_\mathrm{res}\) of the experiment, the Lorentzian contribution represents the linewidth broadening of the electronic band. The total EDC can be represented as a sum over all contributing bands:

\begin{align}
    \mathrm{EDC}_\mathrm{fit}(E) = \sum_n \mathrm{V}_{i}(E).
\end{align}

The index \(n\) runs over the bands contributing to the EDC. The Gaussian width was held fixed, while the Lorentzian widths and peak positions of the Voigt functions were treated as free parameters. To determine the transient of the band energies, the same procedure was applied to the EDCs at each pump--probe delay.

\newpage

\begin{figure}
	\centering
	\includegraphics[width = \textwidth]{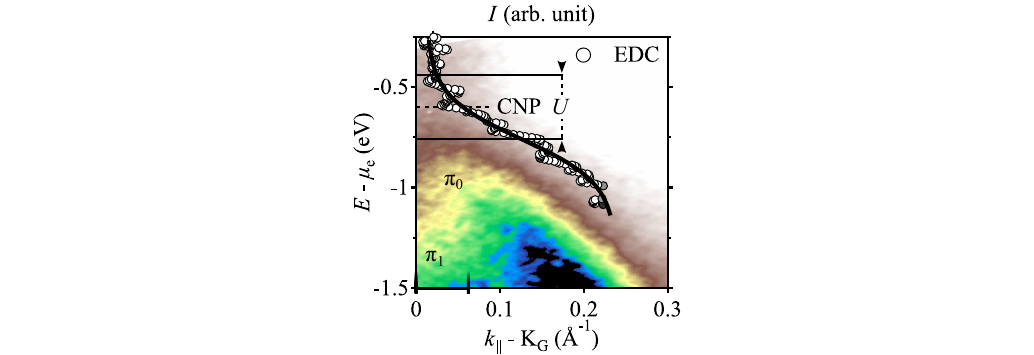} 
	\caption{\textbf{Interlayer potential asymmetry of BLG.} Equilibrium photoemission spectra at \(\mathrm{K}_\mathrm{G}\). The EDC near and within the band-gap region was fitted using a sigmoid function to determine the band-edge energy at \(\mathrm{K}_\mathrm{G}\) of the \(\pi_0\) band. The interlayer potential asymmetry \(U\) was approximated by twice the energy difference between bandedge and charge neutrality point (CNP), corresponding to the doping level of \(\SI{600}{\milli\electronvolt}\) relative to \(\mu_\mathrm{e}\), and the \(\pi_0\)-band edge, i.e., \(U \approx 2 (E_{\pi_0} - E_\mathrm{CNP})\approx \SI{300}{\milli\electronvolt}\).}
	\label{fig:BLG_MLAg_SiC_sup_mat_BLG_band_gap}
\end{figure}

\begin{figure}
	\centering
	\includegraphics[width = \textwidth]{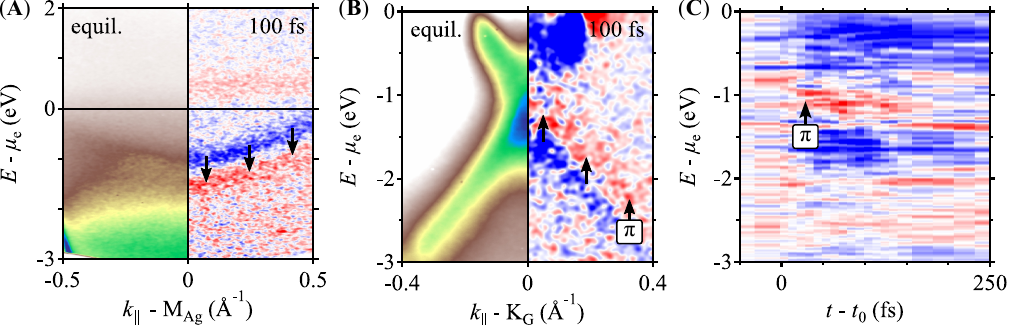} 
	\caption{\textbf{Rigid band shifts in MLAg in BLG/MLAg/6H-SiC(0001) and in MLG in the reference heterostructure MLG/MLAg/6H-SiC(0001).} 
    \textbf{(A)} Equilibrium photoemission spectrum (left) and corresponding differential photoemission spectrum (right) at \(t - t_0 = \SI{100}{\femto\second}\) of MLAg at \(\mathrm{M}_\mathrm{Ag}\). Redistribution of spectral weight below \(\mu_\mathrm{e}\) indicates a rigid downshift of the valence band toward higher binding energies. 
    \textbf{(B)} Equilibrium photoemission spectrum (left) and corresponding differential photoemission spectrum (right) at \(t - t_0 = \SI{100}{\femto\second}\) of MLG at \(\mathrm{K}_\mathrm{G}\). Redistribution of spectral weight below \(\mu_\mathrm{e}\) indicates a rigid upshift of the \(\pi\) band toward lower binding energies, consistent with a transient net negative charging of MLG. The rigid downshift of the valence band in MLAg in this heterostructure shows the same behavior as in BLG/MLAg/6H-SiC(0001). Note that MLG on MLAg does not exhibit a substrate-induced band gap \cite{ArpitarXiv2025}. 
    \textbf{(C)} Transient differential EDCs of MLG at \(\mathrm{K}_\mathrm{G}\). The band-gap region shows a transient redistribution of spectral weight below \(\mu_\mathrm{e}\), indicating a rigid upshift of the \(\pi\) band toward lower binding energies, which decays approximately \(\SI{200}{\femto\second}\) after photoexcitation. There are no indications of band renormalizations as observed in BLG.}
	\label{fig:BLG_MLAg_SiC_sup_mat_MLG_MLAg}
\end{figure}

\begin{figure}
	\centering
	\includegraphics[width = \textwidth]{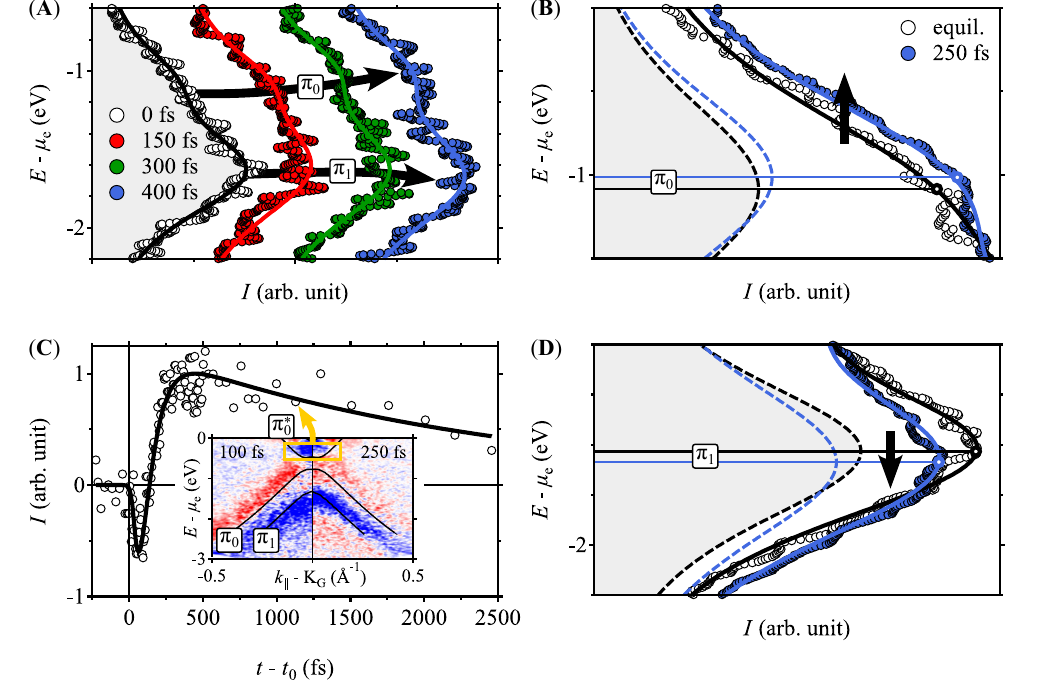} 
	\caption{\textbf{Analysis of the band renormalizations of \(\pi_0\) and \(\pi_1\) band in BLG.} \textbf{(A)} Background-corrected EDCs for different pump--probe delays across the \(\pi_0\) and \(\pi_1\) bands near \(\mathrm{K}_\mathrm{G}\) were analyzed, and the EDCs were fitted using Voigt functions, with the experimental resolution accounted for in the Gaussian width. 
    \textbf{(B, D)} EDCs in equilibrium and at \(t - t_0 = \SI{250}{\femto\second}\) after photoexcitation in expanded view. The solid lines represent the fits to the experimental EDCs, while the dashed lines show the individual contributions of the Voigt functions for \(\pi_0\) and \(\pi_1\) to emphasize the observed band shifts. The \(\pi_0\) band shifts to lower binding energies, whereas the \(\pi_1\) band shifts to higher binding energies, resulting in a clearly resolved splitting. 
    \textbf{(C)} Integrated photoemission intensity as a function of pump-probe delay within the region of interest marked in the inset. The region of interest covers the energy region between \(\mu_\mathrm{e}\) and the CNP. The data show a transient increase in photoemission intensity that is consistent with the dynamic band renormalization of the \(\pi_0\) band and indicates an inverted band renormalization of the \(\pi_0^\ast\) band at \(\mathrm{K}_\mathrm{G}\).}
	\label{fig:BLG_MLAg_SiC_Sup_Bandrenormalization}
\end{figure}

\begin{figure}
	\centering
	\includegraphics[width = \textwidth]{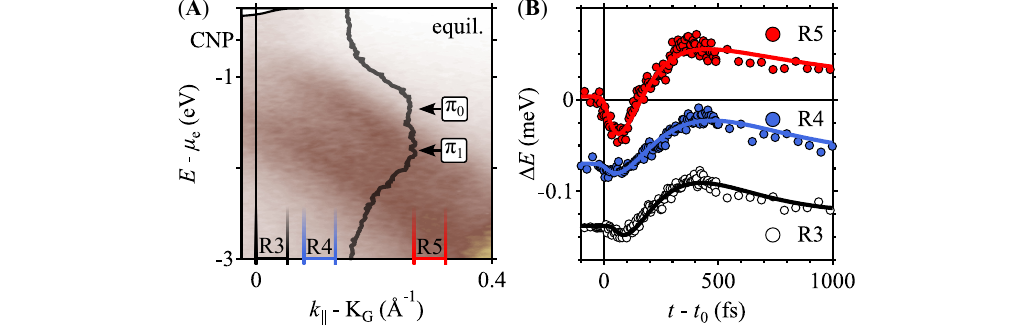} 
	\caption{\textbf{Interlayer coupling in BLG.} \textbf{A} Equilibrium photoemission spectrum of BLG at \(\mathrm{K}_\mathrm{G}\). (\textbf{B}) Changes in the \(\pi_1\) band binding energy as a function of pump--probe delay for the momentum regions of interest R3 -- R5 indicated in panel A. The transient band binding energies are vertically offset for clarity. The transient change in binding energy of the \(\pi_1\) band reflects a transient change of the interlayer coupling \(\gamma_1\).}
	\label{fig:blg_mlag_sic_sup_mat_interlayer_coupling}
\end{figure}

\newpage



\end{document}